\documentclass[%
 reprint,
 amsmath,amssymb,
 aps,
]{revtex4-1}

\usepackage{hyperref}
\usepackage{graphicx}
\usepackage{dcolumn}
\usepackage{bm}
\usepackage[caption=false]{subfig}

\begin{document}

\preprint{APS/123-QED}

\title{Time evolution of intermittency in the passive slider problem}
\author{Tapas Singha} 
\email{tapass@tifrh.res.in}
\author{Mustansir Barma}
\email{barma23@gmail.com}
\affiliation{TIFR Centre for Interdisciplinary Sciences, Tata Institute of Fundamental Research, Gopanpally, Hyderabad-500107, India}
\date{\today}


\date{\today}

\begin{abstract}
How does a steady state with strong intermittency develop in time from an initial state which is statistically random? For passive sliders driven by various fluctuating surfaces, we show that the approach involves an indefinitely growing length scale which governs scaling properties. A simple model of sticky sliders suggests scaling forms for the time-dependent flatness and hyperflatness, both measures of intermittency and these are confirmed numerically for passive sliders driven by a Kardar-Parisi-Zhang surface. Aging properties are studied via a two-time flatness. We predict and verify numerically that the time-dependent flatness is, remarkably, a non-monotonic function of time, with different scaling forms at short and long times. The scaling description remains valid when clustering is more diffuse as for passive sliders evolving through Edwards-Wilkinson driving or under antiadvection, although exponents and scaling functions differ substantially.   
\end{abstract}

\maketitle


Intermittent states in statistical systems driven out of equilibrium are characterized by sudden bursts of activity followed by long periods of stasis. Such states are ubiquitous, and are found in far-flung settings ranging from  turbulent flows \cite{Frisch95} to biological systems on cellular and macroscopic scales \cite{DPR16, SBR13, HA04}. As emphasized by Zeldovich \emph{et al}., a correct average is only a mild virtue when the characteristic structures have intense peaks with short extents in space and time \cite{Zetal}. Evidently, measures which focus on large fluctuations rather than average values, are required in such situations. These are provided by the ratios of moments of time-dependent structure functions; their time dependences quantify the degree of intermittency in such a steady state (ss)\cite{Frisch95}. 

An important question arises: How does intermittency set in, or more precisely, how is an intermittent steady state approached in time, starting from a random initial condition?  It is clearly important to quantify the degree of intermittency or intensity of the peaks as a function of time as also the growth of a two-time correlation while the system approaches the steady state. 

Intermittency in space and time often goes hand in hand with clustering of particles. In this Rapid Communication, we focus on systems of passive particles driven by stochastically evolving fields, under whose influence they cluster. We consider three examples of particles driven by fluctuating surfaces: Kardar-Parisi-Zhang (KPZ) dynamics, with particles moving down slopes in the direction of growth; Edwards-Wilkinson (EW) dynamics, with downward particle motion; and antiadvection (AA) where particles move opposite to the growth direction for a KPZ surface\cite{DK_PRB02,ANMB_PRL05,ANMB_PRE06}. These three modes of driving produce very different degrees of particle clustering. 

Despite these differences, the approach to the steady state is described by scaling in all three cases but with interesting variations. Central to the occurrence of scaling is the existence  of a length scale $\mathcal{L}(t) \sim t^{1/z}$ which grows indefinitely in time and characterizes the sizes of basins which contribute to the growing number of particles in clusters which form at time $t$. The dynamical exponent $z$, which need not be the equal to the dynamical exponent
$z_{\text{surf}}$ for the surface, is different in the three cases, as are the corresponding scaling functions, reflecting different degrees of clustering.  

We also address the question of \emph{aging}, which focuses on correlation functions at two times, as the separation between the times is varied \cite{AB94} during the approach to the steady state. We predict that, remarkably, the time-dependent flatness is a \emph{nonmonotonic} function of the time separation and verify this numerically. 
\begin{figure}[ht!]
\hspace{-1mm}
\centering
\includegraphics[width=6.9cm,height=3.6cm]{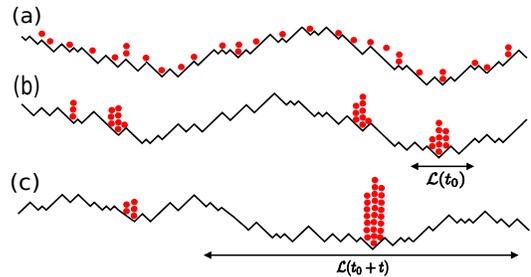}
\caption{A diverging length scale underlies clustering during the approach to an intermittent steady state for passive particles on a fluctuating surface. (a) Initial random placement. (b) By time $t_0$, particles form clusters by gathering together over a length scale of $\mathcal{L}(t_0)$. (c) As time increases to $t_0+t$, the length scale $\mathcal{L}$ and cluster sizes increase.}
\label{Schematic_KPZ}
\end{figure}
 
 The properties of a diffusing passive field which is driven by an extended fluctuating field constitutes the widely studied passive scalar problem \cite{FAL}; the passive slider model (PSM) under study here is a special case. The evolution equations for the PSM are
\begin{eqnarray}\label{eq_KPZ}
\partial h/\partial t &= \nu \nabla^2 h + (\lambda/2) (\nabla h)^2 + \eta(\mathbf{x},t),\\
\label{eq_passive}
\partial \rho/\partial t &=  \nabla \cdot [\kappa \nabla \rho + a (\rho \nabla h) + \xi (\mathbf{x},t)],
\end{eqnarray}
where $\eta (\mathbf{x},t)$ and $\xi (\mathbf{x},t)$ are spatially and temporally uncorrelated Gaussian white noises with zero mean and unit variance. Equation (\ref{eq_KPZ}) is the KPZ equation for the evolution of the height $h(x,t)$ of a growing interface, where $\nu$ and $\lambda$, respectively, are the surface tension and growth-related nonlinear coefficient \cite{KPZ,Medina}. Equation \ref{eq_passive} describes the motion of passive particles which tend to follow the local slope of the surface in addition to being subjected to noise. On substituting $\nabla h=-\mathbf{u}$ and setting $\lambda=1$, Eq.\ref{eq_KPZ} maps onto the noisy Burgers equation in the zero-vorticity sector \cite{Medina, BEC}. On setting $\lambda=0$, Eq.\ref{eq_KPZ} reduces to the EW model \cite{EW}  which describes a fluctuating interface in equilibrium. 


We study a discrete lattice model \cite{ANMB_PRL05,ANMB_PRE06} whose behavior at long length and time scales is described by Eqs.\ref{eq_KPZ} and \ref{eq_passive}. In one dimension (1D), individual bonds of the lattice can take on one of two orientations $/$ or $\backslash$, and the time evolution of the surface stochastically transforms local hills $/$$\backslash$ into local valleys $\backslash$$/$ at rate $u_1$ and local valleys $\backslash$$/$ into local hills $/$$\backslash$ at rate $u_2$. The EW case corresponds to $u_1=u_2$, whereas $u_1$$>$$u_2$ leads to KPZ dynamics at large scales of length and time. The associated dynamical exponent is $z_{\text{surf}}$, which has the values $\frac{3}{2}$ and $2$ for KPZ and EW dynamics, respectively. Passive sliders are represented as noninteracting particles which reside on the sites between bonds. A randomly chosen particle slides down if it is atop a bond; if atop a local hill $/$$\backslash$, it slides down one of the two bonds with equal likelihood, and if in a valley $\backslash$$/$, it does not move. In the numerical simulations reported below, we take the number of particles $N$ to be the number of sites $L$ and attempt updates of randomly chosen particles and surface bonds at equal rates. Antiadvection corresponds to the case of $u_1$$<$$u_2$ in which case the direction of particle sliding is opposite to that of surface growth.

The strong tendency of particles to cluster together (Fig.\ref{Schematic_KPZ}) in the KPZ-PSM is well established, both without \cite{BP94,CSC}, and with diffusion \cite{DK_PRL00,DK_PRB02}. A similar conclusion was reached, by showing that the overlap of two-particle trajectories has an anomalous two-peaked distribution \cite{US02}. Furthermore, if $n_i$ is the number of particles on site $i$, the correlation function $C(r,L) \equiv \langle n_i n_{i+r}\rangle$ in a system of size $L$ was found to follow 
\begin{equation}
C(r,L) = c_1 L \ \delta_{r,0} + c_2 L^{-1/2}f(r/L)
\label{Eq_correlation}
\end{equation}
where $c_1$ and $c_2$ are the constants and $f(y)$ decays as $y^{-3/2}$ \cite{ANMB_PRE06}, indicating clustering with a size-dependent power-law tail. Passive sliders on an EW interface also exhibit clustering, but of a more diffuse variety: $C(r,L)$ does not have a $\delta$ function part and decays as $(r/L)^{-2/3}$ \cite{ANMB_PRE06}. In the case of AA, clustering is yet more diffuse, with $C(r,L)$ decaying as $(r/L)^{-1/3}$.

Clustering of inertial particles in compressible \cite{BFF} and incompressible \cite{FAL} turbulent flows lead to the growth of density fluctuations. Moments of the density were found to grow exponentially, but are diffusion limited. 
 
 Intermittency is related to clustering as the formation of a large cluster depletes particle density in the extended neighborhood. As a function of time, the resulting poorly populated regions show low activity, punctuated by high signals only when the cluster visits that region. To study this quantitatively, define the number of particles $N_l(t)$  at time $t$ in a stretch of $l$ sites which is a small  fraction of the overall size $L$ \cite{important}. We study the structure functions
\begin{equation}
S_q(t_0,t,l) = \langle [N_l(t_0+t)-N_l(t_0)]^{q} \rangle
\label{struc}
\end{equation}
for $q=2$, $4$, and $6$. Direct measures of the degree of intermittency are provided by the flatness $\kappa_4(t_0,t,l)$ and hyperflatness $\kappa_6(t_0,t,l)$, defined as
\begin{equation}
 \kappa_4(t_0,t,l)=\frac{S_4(t_0,t,l)}{S^2_2(t_0,t,l)},    \  \  \   \   \ \kappa_6(t_0,t,l)=\frac{S_6(t_0,t,l)}{S^3_2(t_0,t,l).} 
  \label{flat_steady}
\end{equation}
Sudden changes in $N_l(t)$ are well captured by $S_q(t_0, t,l)$, and result in diverging $\kappa_4$ and $\kappa_6$ in the limit $t/L^z \rightarrow 0$  \cite{Frisch95}. This indicates intermittency. 

It is useful to recall how a single passive particle is advected; its displacement grows as $r(t) \sim t^{1/z}$ where $1/z =\frac{2}{3}$ for KPZ driving, but $1/z \simeq 0.56$ for EW driving whereas for the case of AA, we find $1/z \simeq 0.57$ with equal update rates for the surface and particles \cite{Manoj, Huveneers}. Note that $z$ coincides with $z_{\text{surf}}$ only for the KPZ case, indicating that particle motion is militated purely by the surface dynamics only in that case.

In the remainder of the Rapid Communication, we first discuss the case of KPZ driving as the clustering properties of particles are simpler. This is followed by a discussion of intermittency in the EW and AA cases, highlighting the differences which arise from the diffuse nature of clustering in these cases. 

\begin{figure}[ht]
\vspace{3mm}
\hspace{-3mm}
\centering
\includegraphics[width=8.1cm,height=6cm]{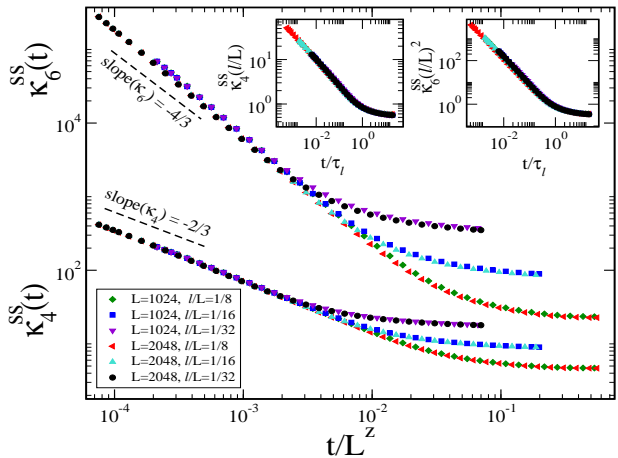}
\caption{Steady state with KPZ driving: $\kappa^{\text{ss}}_4(t)$ and $\kappa^{\text{ss}}_6(t)$ for different fractions $l/L$. Inset: Scaling collapse is obtained with $t$ scaled by $\tau_l$ and $\kappa^{\text{ss}}_4$ and $\kappa^{\text{ss}}_6$ by powers of $(l/L).$ The divergences at low argument indicate intermittency.}
\label{SteadyState_KPZ}
\end{figure}

We begin by providing evidence that the steady state of the KPZ-PSM exhibits intermittency. To this end, we monitor the $t$ dependence of the structure functions $S^{\text{ss}}_q$ and their ratios $\kappa^{\text{ss}}_4$ and $\kappa^{\text{ss}}_6$, obtained by taking $t_0 \gg L^{z}$ in Eqs.\ref{struc} and \ref{flat_steady}. Our numerical results show that for a fixed value of the fraction  $l/L$ both $\kappa^{\text{ss}}_4$ and $\kappa^{\text{ss}}_6$  exhibit a scaling collapse when plotted against $t/L^{z}$ (Fig.\ref{SteadyState_KPZ}). Particles cluster strongly in valleys \cite{CSC,US02}. Thus the particle displacement $\sim t^{1/z}$ with $z=\frac{3}{2}$. As $t/L^{z}  \rightarrow 0$, the data in Fig.\ref{SteadyState_KPZ} suggests a power-law divergence of $\kappa^{\text{ss}}_4$ and $\kappa^{\text{ss}}_6$ with exponents $\simeq -\frac{2}{3}$ and $\simeq -\frac{4}{3}$, respectively. As $t/L^z  \rightarrow \infty$, we find $\kappa^{\text{ss}}_4$ and $\kappa^{\text{ss}}_6$ saturate at values which depend on $l/L$. However, as evident from the insets in Fig.\ref{SteadyState_KPZ}, a scaling collapse of all curves for $q=4$ and $6$ is obtained by rescaling $t$ by $\tau_l$, and $\kappa^{\text{ss}}_4$ and $\kappa^{\text{ss}}_6$ by powers of $(l/L)$, where $\tau_l \sim l^{z}$. These behaviors are summarized by the scaling form  
\begin{eqnarray}
\kappa^{ss}_q \sim (L/l)^{\gamma_q}  F_q(t/\tau_{l}),  
\label{eq:SS_scaling}
\end{eqnarray}
with $F_q(y) \sim y^{-\gamma_q/z}$ as $y \rightarrow 0$ and $F_q \rightarrow \text{const}.$ as $y \rightarrow \infty.$ The numerical results yield $\gamma_4 \simeq 1$ and $\gamma_6 \simeq 2$ (Fig.\ref{SteadyState_KPZ}). It is interesting to compare these results with the adiabatic approximation \cite{AdiaApp,BMP}. We find that the approximation performs rather poorly; although the adiabatic density does exhibit intermittency, the associated powers are substantially lower than for the PSM.  

To get an insight into the occurrence of scaling, we define a sticky slider model (SSM) in which particles driven by the surface undergo irreversible aggregation and do not separate, once they are on the same site. We emphasize that our purpose in invoking the SSM is only to suggest scaling forms in steady state, and during the approach to it. Despite the fact that the cluster can break up and reconstitute in the models under consideration, the scaling forms are found to remain valid. As the SSM promotes strong clustering it provides a valuable benchmark to compare with for various types of driving.

Evidently, in the steady state of the SSM, a single aggregate $A$ with $N$ particles slides stochastically on a $\text{1D}$ fluctuating KPZ surface of size $L$. In time $t$, $A$ can travel typically a distance $r(t)\sim t^{1/z}$. Its motion is identical to that of a single particle driven by a fluctuating interface \cite{DK_PRB02}. First consider the steady-state structure functions, $S^{ss}_q$, in a stretch $l$ in the small time limit, i.e., $t \ll \tau_l$. Let the locations of $A$ at time $t_0$ and $t_0+t$ be $R_0$ and $R$, respectively. The probability that $R_0$ is inside the stretch $l$ is $l/L$. Given this, the probability of $R$ falling outside $l$ is of the order of $p_1=r(t)/L$. Similarly, given $R_0$ outside $l$, the probability of $R$ falling inside $l$ is $p_2=(1-\frac{l}{L})\frac{r(t)}{L}$. Within the SSM, only these two options give nonzero contributions to the structure functions. We find $S^{ss}_2=p_1 N^2+p_2 N^2=(r(t)/L)(2-l/L) N^2$ and $S^{ss}_4(t)=p_1 N^4+p_2 N^4=(r(t)/L)(2-l/L)N^4$, respectively. Since we consider $N=L$ in our simulation, the flatness is $\kappa^{\text{ss}}_4=\frac{L}{r(t)} \frac{1}{(2-l/L)}\approx L/r(t)$ for large $L$. Similarly we find $\kappa^{ss}_6 \approx (L/r(t))^2$. On the other hand, when $t \gg \tau_l$, the probability of $R$ being  inside or outside stretch $l$ is independent of $R_{0}$ leading to $\kappa_4 \approx (L/l)$ and $\kappa_6 \approx (L/l)^2$. Moreover, the SSM suggests that the exponent $\gamma_4=1/z$. 

We see that numerical results discussed above with KPZ driving agree well with the SSM predictions. This level of agreement is not obvious \emph{a priori}, as clustering in the KPZ case is not as intense as for the SSM, as indicated by 
the fact that the correlation function (Eq. \ref{Eq_correlation}) has a power-law tail, as does the distribution of particle number on a given site \cite{ANMB_PRE06}.


Having established the existence of an intermittent steady state in the KPZ-PSM, we now address the central question raised in the Rapid Communication, namely how this steady state is approached, starting with an initial state with randomly placed particles.

We take a cue from the theory of phase ordering, where a successful scaling theory describes the growth of correlation functions in the coarsening regime \cite{AB94}. Underlying the scaling is a growing length scale $\mathcal{L}(t) \sim t^{1/z_{\text{coarsen}}}$ which describes the linear size of a growing droplet. As demonstrated below, a scaling description based on a growing length scale is valid in our problem as well. The difference is that $\mathcal{L}(t)$ denotes not the size of a cluster, but rather the size of the basin from which particles congregate and form  clusters, up to time $t$ (Fig. \ref{Schematic_KPZ}). Since a single particle moves a distance of $r(t)\sim t^{1/z}$ as discussed above, it is plausible that a similar law will describe the distance moved by congregating particles in time $t$ so that we expect $\mathcal{L}(t) \sim t^{1/z}$. Moreover, since the particles are initially distributed with a uniform density $\rho$, the number of particles in the clusters formed at time $t$ is $\rho \mathcal{L}(t)$, where $\rho$ has been taken to be unity in our simulations.

\begin{figure}[ht]
\centering
\includegraphics[width=7cm,height=4.6cm]{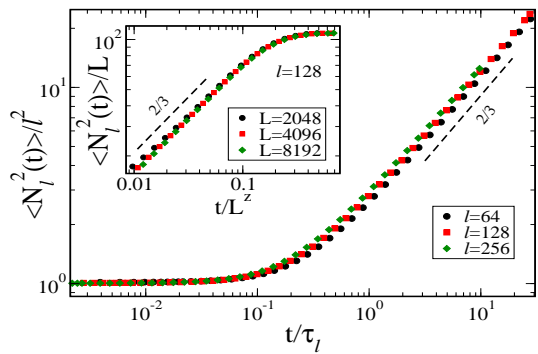}
\caption{Coarsening with KPZ driving: The increase in $\langle N^2_l(t)\rangle$ reflects the growth of cluster sizes in time. The main plot and inset, respectively, show the scaling with $l$ and $L$.} 
\label{2nd_moment}
\end{figure}

Evidence for this is obtained by monitoring the mean squared number $\langle N_l^2(t)\rangle$ of particles in a stretch of $l$ successive sites in a KPZ system of size $L$.  Figure \ref{2nd_moment} shows that (i) starting from a constant value, (ii) $\langle N_l^2(t)\rangle$ increases as a power of time and (iii) finally saturates (the inset). The SSM exhibits these three regimes. (i) For $\mathcal{L}(t)< l$, we have $\langle N_l^2(t)\rangle \approx  l^2$, as clustering on length scales smaller than $l$ does not affect $N_l$. (ii) When  $ l < \mathcal{L}(t) < L$, there is a growth of cluster sizes, as a lateral basin of size $\mathcal{L}(t)$ typically holds a cluster containing
 $\mathcal{L}(t)$ particles. The probability that this cluster is present in the $l$-stretch is $l/\mathcal{L}(t)$, leading to $\langle N_l^2(t)\rangle \sim l \mathcal{L}(t)$. (iii) Finally at very long times $t\gg L^{z}$, the cluster size saturates at $L$, leading to $ \langle N_l^2(t)\rangle \sim l L$. Numerical simulations verify these predictions for the coarsening regime.  Figure \ref{2nd_moment} shows the scaling collapse for different stretch length $l$ for fixed $L$ whereas the inset shows the collapse for different $L$ when $l$ is held fixed. 

To study the growth of temporal intermittency in this regime, we define the structure factors $S_q^c(t,l)$ and their ratios $\kappa^c_4(t_0,t,l)$ and $\kappa^c_6(t_0,t,l)$ by setting $t_0 = 0$ in Eqs. \ref{struc} and \ref{flat_steady} and replacing $N_l(t_0)$ by its average value $N l/L$. 

\begin{figure}[ht]
\vspace{3mm}
\hspace{-5mm}
\includegraphics[width=7.5cm,height=5cm]{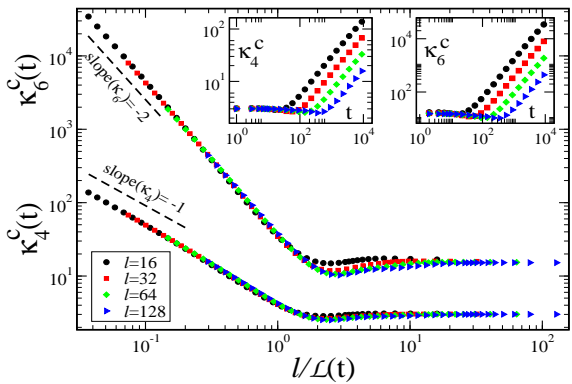}
\caption{Coarsening regime with KPZ driving: $\kappa^c_4(t)$ and $\kappa^c_6(t)$ for different stretch lengths increase in time (the inset). In the main plot, $l$ is scaled by $\mathcal{L}(t)$ leading to data collapse.}
\label{Coarsening}
\end{figure}

Before discussing our numerical results for the PSM, let us examine the scaling form and exponents predicted by the SSM. Within the SSM, $\mathcal{L}(t)$ denotes the typical separation between clusters. Each cluster typically holds $\mathcal{L}(t)$ particles (recall that the particle density is assumed to be unity). Since for KPZ-driven surfaces, valleys tend to coalesce carrying passive sliders with them \cite{CSC}, we expect $\mathcal{L}(t) \sim t^{1/z}$ where $z=\frac{3}{2}$. For very long times $t \gg \tau_l$, the length scale $\mathcal{L}(t)$ greatly exceeds the stretch length $l$. The probability that the aggregate lies within $l$ is then $l/\mathcal{L}(t)$ implying $S_q^c \sim [l/\mathcal{L}(t)] [\mathcal{L}(t)]^q$, leading to $\kappa_4^c = \mathcal{L}(t)/l$ and $\kappa_6^c = [\mathcal{L}(t)/ l]^2$. Thus the SSM predicts that flatness and hyperflatness both diverge as $t \rightarrow \infty$. By contrast, if the time $t \ll \tau_l$, both $S_2^c$  and $S_4^c$ assume constant values. Both behaviors are subsumed in the scaling form:
\begin{equation}
\kappa_q^c \sim G_q^c [l/\mathcal{L}(t)]
\label{eq:Coar_scaling}
\end{equation}
with $G_q^c(y) \sim y^{-\psi_q}$ as $y\rightarrow 0$ and $G_q^c(y) \rightarrow \text{const}.$ as $y \rightarrow \infty$. For the SSM, the exponents $\psi_q$ take on the values 1 and 2 for $q=1$ and 2, respectively. Numerical simulations of the PSM with KPZ driving show scaling, as predicted by the SSM (Fig.\ref{Coarsening}). The data for the flatness and hyperflatness collapse when $l$ is scaled with $\mathcal{L}(t)$ and indicate a power law with the exponents close to the SSM result.

\begin{figure}[ht!]
\vspace{3mm}
\hspace{-0.6cm}
\centering
\includegraphics[width=8cm,height=7.5cm]{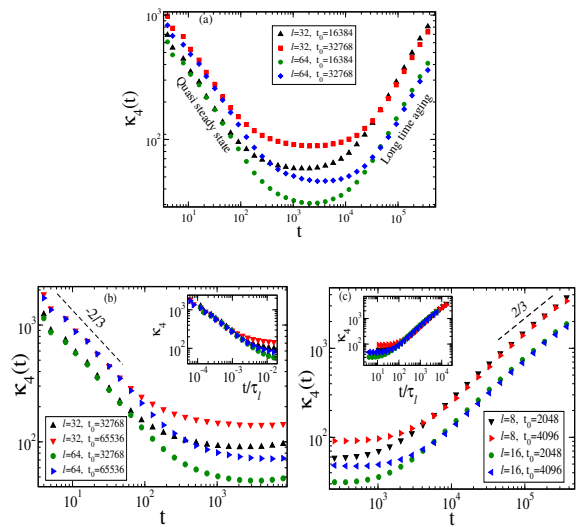}
\caption{Aging with KPZ driving: (a) Flatness $\kappa_4(t)$, of two-time particle number fluctuations, is nonmonotonic as a function of time difference $t$. The left and right wings of the curves scale differently. (b) The quasi steady-state regime, for $t \ll \tau_l$, with data collapse (the inset) when $t$ is scaled by $t_0$. (c) The long time aging (LTA) regime for $t \gg t_0$ with data collapse (the inset) when $t$ is scaled by $\tau_l$.}
\label{Aging}
\end{figure}

Within the coarsening regime, it is also interesting to investigate the \emph{aging} properties. For phase ordering kinetics, one customarily  investigates the two-time auto-correlation function \cite{AB94}, and this has been studied for KPZ surfaces as well \cite{Ramasco,CB06,Bustingorry,HNP12}. In our case, we are interested in the time evolution of intermittency correlations, and so characterize the history-dependent time evolution of the signal via $S_q(t_0,t,l)$ (Eq. \ref{struc}) and $\kappa_q(t_0,t,l)$ (Eq. \ref{flat_steady}) where $t_0$ is the waiting time and $t$ is the further time difference. The most interesting behavior is found for $t_0 \gg \tau_l$, and we restrict the discussion to this regime. Figure \ref{Aging}a shows that for the PSM, $\kappa_4$ is a nonmonotonic function of time difference $t$.   
  
To understand the nonmonotonicity, we invoke the SSM to estimate $S_q(t_0,t,l)$. For $t \ll t_0$, both the linear extent of a typical valley and the number of particles in a single aggregate in the valley are of order $\mathcal{L}(t_0)$ (Fig.\ref{Schematic_KPZ}). In addition, the cluster can be anywhere within the valley of size $\mathcal{L}(t_0)$, reminiscent of steady state, where the aggregate can be anywhere within system size $L$. In this quasi-steady state (QSS) regime, we follow the probabilistic argument for the SSM in steady state for the calculation of structure functions, replacing $L$ by $\mathcal{L}(t_0)$. We find $S_2(t_0,t,l)=r(t) \mathcal{L}(t_0)$ and $S_4(t_0,t,l)=r(t) \mathcal{L}^3(t_0)$, respectively, where $r(t)\sim t^{1/z}$ is the typical distance traveled by the aggregate. Hence we obtain
\begin{equation}
\kappa_4 \sim (t_0/t)^{1/z}  \ \ \ \ \  t \ll \tau_l
\label{QSS}
\end{equation}
in the limit $t_0 \gg \tau_l \gg t$,

In the LTA regime $\tau_l \ll t_0 \ll t \ll L^z$, the location $R_{0}$ (at time $t_0$) is uncorrelated with $R$ (at time $t_0+t$). Thus, we find $ S_q(t_0,t,l) \approx \ l \ [\mathcal{L}(t_0+t)]^{q-1}$. Dropping $t_0$ in the comparison to $t$, we have
\begin{equation}
\kappa_4 \sim (t/\tau_l)^{1/z}  \ \ \ \  \  t \gg t_0  . 
\label{flat_Aging}  
\end{equation}

Finally, let us discuss the intermediate time regime $\tau_l \ll t \ll t_0$. On setting $t=\tau_l$ in the QSS expression Eq.\ref{QSS}, we find $\kappa_4 \sim (t_0/\tau_l)^{1/z}$. This coincides with the value obtained by setting $t=t_0$ in Eq.\ref{flat_Aging} which describes the LTA regime. We conclude that $\kappa_4$ does not change much at intermediate times, although a mild variation cannot be ruled out.
These predictions agree well with the numerical data for the PSM. 

In Fig.\ref{Aging}c, the data of flatness in the LTA regime for different 
$t_0'$s and $l'$s are shown. Data collapse is observed when $t$ is scaled by $\tau_l$, with a power-law with exponent $\simeq \frac{2}{3}$ as predicted by Eq.\ref{flat_Aging} for the SSM. In the QSS regime, by contrast, the data for $\kappa_4(t_0,t,l)$ for different stretches of $l$ and $t_0$ are found to collapse when $t$ is scaled by $t_0$ as shown in Fig.\ref{Aging}b. The scaling form and exponent ($\simeq -\frac{2}{3}$) for $\kappa_4(t_0,t,l)$ match well with the SSM prediction Eq.\ref{QSS}. 
\begin{figure}[ht!]
 \centering
 \includegraphics[width=7.5cm,height=7cm]{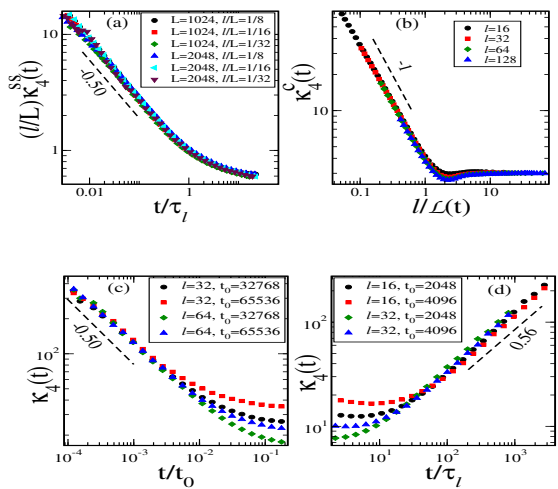}
\caption{Results with EW driving: (a) Scaled flatness with scaled time for different $l'$s in the steady state. (b) Flatness with scaled $l$ in the coarsening regime. (c) Flatness with $t$ for different $l$ and different $t_0'$s when $t$ is scaled by $t_0$ in the quasi-steady state regime. (d) Flatness with $t$ for different $l'$s and $t_0's$ when $t$ is scaled by $\tau_l$ in the long time aging regime.}
\label{EWALL}
\end{figure}

For the SSM, simple physical arguments explain the diverging flatness at small and large $t$, in terms of rare events.  For small $t$, the event that contributes to $S_q$ (Eq. \ref{struc}) is the movement of the cluster across the boundary of the $l$-stretch in time $t$, associated with a crossing probability $p^{\text{cross}} \sim r(t)/ \mathcal{L}(t_0)$. At large $t$, the condition is that the stretch must contain the cluster at time $t_0+t$; the corresponding occupation probability is $p^{occ} \sim l / \mathcal{L}(t_0+t)$. Thus the flatness, which is given by the inverse of these probabilities, diverges both as $t \rightarrow 0$ and as $t \rightarrow \infty$.

The question arises whether the scaling descriptions of the intermittent steady state and the approach to it remain valid even when clustering is significantly less intense than for KPZ driving. To answer this, we numerically studied the PSM with EW dynamics, as well as antiadvection \cite{ANMB_PRE06, DK_PRB02}, where clustering is markedly weaker \cite{ANMB_PRE06}. Results for $\kappa_4$ for the EW model are shown in Fig.\ref{EWALL}. The scaling forms of Eqs. \ref{flat_steady} and \ref{eq:SS_scaling} for steady state dynamics and coarsening remain valid, as seen in Figs.\ref{EWALL}a and \ref{EWALL}b, with $1/z \simeq 0.56$ and new scaling functions $F_4$ and $G_4$.  Also, in the aging regime, $\kappa_4$ varies nonmonotonically with time, and Figs. \ref{EWALL}c and \ref{EWALL}d show scaling collapses in the QSS and LTA regimes.  Likewise, for the PSM with AA dynamics, we have verified that Eqs. \ref{flat_steady} and \ref{eq:SS_scaling} hold, and that a similar description holds in the aging regime. Not surprisingly, the scaling functions for the PSMs for EW and AA dynamics differ substantially from those for the corresponding SSMs. 

We conclude that for a broad class of systems with an intermittent steady state which exhibits particle clustering, whether strong or relatively weak, the scaling description of the dynamics of approach to the steady state should remain valid, with a concomitant diverging length scale of $\mathcal{L}(t)\sim t^{1/z}$. Differences between systems would manifest themselves in the value of the dynamical exponent $z$, and the forms of the scaling functions. We expect this description to work for a number of systems which exhibit clustering such as inelastically colliding particles \cite{SDR}, aggregating clusters in open systems \cite{SB14}, particles driven by a spatially correlated Gaussian field leading to path coalescence \cite{DE85,WM03} and the zero-range process \cite{Evans}, along with the physical systems they describe.

\vspace{4mm}

We thank P. Perlekar, F. Huveneers and T. Sadhu for useful discussions.


\begin{thebibliography}{10}
\bibitem{important}
It is important to monitor a stretch, as opposed to a single site, so that particles on close-by sites are captured.
\bibitem{AdiaApp}
In the adiabatic approximation, the particles are taken to establish the Boltzmann distribution with respect to the height field, and to follow the latter as it evolves in time.
\end{thebibliography}
\end{document}